# One cut-point phase-type distributions in Reliability. An application to Resistive Random Access Memories

Christian Acal[1], Juan E. Ruiz-Castro[1,*], David Maldonado[2] and Juan B. Roldán[2]

[1] Department of Statistics and O.R. and IMAG, University of Granada, 18071 Granada, Spain; chracal@ugr.es (C.A.); jeloy@ugr.es (J.E.R.C.)
[2] Department of Electronics and Computing Technology, University of Granada, 18071 Granada, Spain; dmaldonado@ugr.es (D.M.); jroldan@ugr.es (J.B.R.)
* Correspondence: jeloy@ugr.es (J.E.R.C.)

**Abstract:** A new probability distribution to study lifetime data in reliability is introduced in this paper. This one is a first approach to a non-homogeneous phase-type distribution. It is built by considering one cut-point in the non-negative semi-line of a phase-type distribution. The density function is defined and the main measures associated, such as the reliability function, hazard rate, cumulative hazard rate and the characteristic function are also worked out. This new class of distributions enables to decrease the number of parameter in the estimate when inference is considered. Besides, the likelihood distribution is built to estimate the model parameters by maximum likelihood. Several applications by considering Resistive Random Access Memories compare the adjustment when phase type distributions and one cut-point phase-type distributions are considered. The developed methodology has been computationally implemented in R-cran.

## 1. Introduction

Parametric probability distributions are used in several specific fields such as reliability of complex systems and different industrial applications, among them, electronics. The study of reliability is an essential tool at the industry, mostly when products are made of a high number of different components. These probability distributions are assumed to be fully known and the corresponding properties are analysed in depth. A specific statistical inference is carried out to estimate the parameters of the distribution from a process data. In this context, maximum likelihood is one of the most considered methodologies used thanks to the good asymptotic properties of the estimators. Care must be taken to diagnose the underlying assumptions associated with the assumed specific parametric distribution to ensure a reasonable fit with the empirical data.

A detailed review of the most used basic parametric probability distributions in reliability can be seen in [1-3]. Key text books in reliability and related topics tend to present parametric probability distributions in detail, see e.g. [4-5]. Some of the most commonly used parametric probability distributions are the exponential, Weibull and log-normal distribution. [6] proposed a new approach based on different Weibull distributions for modeling the lifetime data in systems whose components randomly fail. Recently, in [7] is introduced a new lifetime distribution by considering a series system such that the components are log-normal and Weibull distributed. This lifetime model proposes a new flexible lifetime distribution that can be used to model lifetime data in a wider class of reliability problems. The parameters are estimated by considering maximum likelihood.

One of the most interesting distribution class, used to model multi-state reliability systems, is the phase-type distribution (PH). This class of distributions were introduced and described in detail by [8]. The good properties of these distributions enable to model complex systems in an algorithmic and computational form. Besides, this class of distributions is dense in the set of non-negative probability distributions [9]. Thus, any non-negative distribution can be approximated as much as desired through a PH. In relation to the parameter estimation of PH probability distributions by maximum likelihood, a recurring method called EM algorithm is often employed [10]. Recently, [11-12] showed that PH distributions explain the behaviour of Resistive Random Access Memories (RRAM) better than Weibull distributions, as had been considered up to now.

One of the main problems when PH distributions are considered to be fitted to a data set is the number of parameters. The number of phases of the PH distribution increases as the data set variability rises, and then, the number of parameters to be estimated could be very high [13]. Another consideration is when the empirical distribution has multiple modes or the tail is heavy. Again, the number of phases is high in these situations. As models for heavy tailed distributions, a new class of infinite-dimensional PH distributions with finitely many parameters was proposed in [14], meanwhile [15] provided a new solution by introducing time-inhomogeneity in the Markov jump process underlying the construction of the PH distribution. The authors transform PH distributions into heavy-tailed ones, but rather than transforming the PH distribution directly, they transform the time scales of each state of the underlying Markov process. They introduce the matrix-Pareto distribution and they fit this one to a real dataset by considering an Erlang PH structure. A different option is to consider simpler structures that are more attractive both for fitting and simulation. In this line, multiple analysis have being developed by using specific structures for the PH distribution such as Coxian or Erlang distributions [16-17], whose number of parameters to be estimated is lower [18-19]. However, even using these specific structures, in which the dimension of estimation is smaller, there are situations where the adjustment is not all suitable.

Regarding these problems, as a kind of example/motivation, in [13] forming voltage distributions for different RRAMs were fitted by using PH distributions. Several cases did not provide optimal results when a PH distribution was fitted. In particular, for the dataset related to the devices with dielectric HfAlO and subjected to a temperature of 80ºC (see Table 1 and 2 therein), the best PH-distribution was achieved with 128 phases. Even being the number of phases really high, the fitting is not as satisfactory as one could expect. Another example can be seen in [11] with the modelling of reset voltages (voltage in which the RRAMs' conductive filament is broken). Here, the optimum number of states is only 15, but as it is displayed in Figure 1, empirical and theoretical hazard rate show a non-accurate fitting. Note that the empirical hazard rate has been obtained by means of "*muhaz*" package available in R-cran. By default, this function estimates the hazard function using the Epanechnikov kernel and the local method for bandwidth functions, in which the optimal one is worked out at a grid point by minimizing the local mean squared error [20].

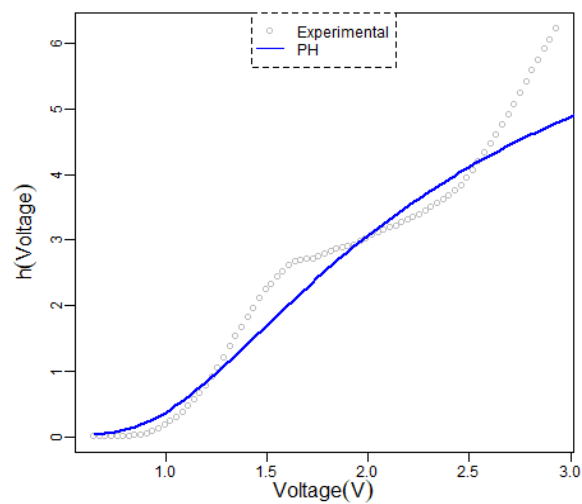

**Figure 1**: Statistical PH fit of the Hazard Rate of reset voltages collected from Ni/HfO$_2$/Si RRAM samples.

The motivation of this paper is to provide a new tool that improves the quality of the fitting for similar scenarios. For that purpose, a new distribution called one cut-point PH distribution is introduced. This distribution is a first approach to non-homogeneous PH distributions. The main measures associated to this class can be expressed in an algorithmic and computational form, as the original PH distributions. Also, this class will enable to get adjustments with a lower number of parameter when the embedded distribution presents some of the features aforementioned. This new distribution is defined and the main measures are worked out. In addition, the likelihood distribution is built to estimate the model parameters. The one cut-point PH distribution is applied to analyse the behaviour of a conductive filament (CF) in RRAMs with filamentary charge transport, according to several characteristics [13,21].

Resistive Random Access Memories are under scrutiny both at the academia and the industry [21] due to their potential in applications related to non-volatile memories, neuromorphic computing and hardware cryptography [22]; they are used in entropy sources for the implementation of physical unclonable functions and random number generators. The operation of most of these devices is filamentary, it consists of the formation and destruction of a conductive filament that changes the device resistance. The filament can be modified many times in a process known as resistive switching, which is characterized by the constant random creation (set process) and rupture (reset process) of the conductive filament. The CF, controlled by means of electric signals, allows both digital and analog operation in different applications with very good perspectives since these devices can be easily scaled to keep up with the continuous miniaturization race that it is taking place in the industry [21].

The paper is organized as follows. Section 2 motivates the new distribution from a real problem and the PH distributions class is remembered. The new distribution and features are given in section 3. Several real applications show the versatility of this new distribution in section 4. Conclusions are given in section 5.

## 2. Introducing the one cut-point PH distribution

We assume a device with several internal transient unobservable states, $S = \{1, 2, ..., n\}$. The performance of the device goes through these stages up to reach an event, state $n+1$. If the internal behaviour of the device is Markovian, then the time up to this event is phase-type distributed [8].

The generator of the Markov process that governs the internal behaviour is a $Q$-matrix with order $n+1$ where the last state is an absorbing state. This generator, with order $n+1$ x $n+1$, can be expressed by blocks as follows

$$\mathbf{Q} = \left( \begin{array}{c|c} \mathbf{T} & \mathbf{T}^0 \\ \hline \mathbf{0} & 0 \end{array} \right),$$

where the matrix $\mathbf{T}$, order $n$ x $n$, contains the transition intensities between transient states and the column vector $\mathbf{T}^0$, order $n$ x 1, the transition intensities from a transient state up to the absorbing event. Matrix $\mathbf{T}$ is a non-singular matrix given the embedded Markov structure. Throughout the paper, given a matrix $\mathbf{A}$, the column vector $\mathbf{A}^0$ is given by $\mathbf{A}^0 = -\mathbf{A}\mathbf{e}$, being $\mathbf{e}$ a column vector of ones with appropriate order. If $\mathbf{A}$ is a non-singular matrix then $\mathbf{A}^{-1}\mathbf{A}^0 = -\mathbf{e}$ and $\mathbf{A}^{-2}\mathbf{A}^0 = -\mathbf{A}^{-1}\mathbf{e}$.

Then, it is well-known that the time up to the absorption, $X$, when the initial distribution for the transient states is the row vector $\boldsymbol{\alpha}$, is PH distributed with representation $(\boldsymbol{\alpha}, \mathbf{T})$. The reliability function of a PH distribution with representation $(\boldsymbol{\alpha}, \mathbf{T})$ is given by $R(x) = P(X > x) = \boldsymbol{\alpha} e^{\mathbf{T}x} \mathbf{e}$. It can be interpreted as follows. The element $(i, j)$ of the matrix $e^{\mathbf{T}x}$ is the probability of being in state $j$ at time $x$ given that initially it is in state $i$. Therefore, $\boldsymbol{\alpha} e^{\mathbf{T}x} \mathbf{e}$ is the probability of being in a transient state at time $x$.

As it has been mentioned in the introduction, one of the main problems when PH distributions are considered is the order and the internal structure of the matrix $\mathbf{T}$ and therefore, the number of parameters to estimate. The problem described in the introduction from [11] and [13] can be provoked because the internal behaviour of the states is not the same over time, that is, the internal behaviour is non-homogeneous. From Figure 1, it can be observed that the data set has different performances. A first approach is to consider one cut-point by partitioning the real line in two semi-lines. We can consider that the internal behaviour of the device passes across $n$ states but the intensities are not the same in each interval. If the cut-point is denoted by $a$, which can be estimated, the transition intensities are given, for period one and two, from matrices $\mathbf{T}_1$ and $\mathbf{T}_2$ respectively. Figure 2 shows the situation for the motivation example from the hazard rate function.

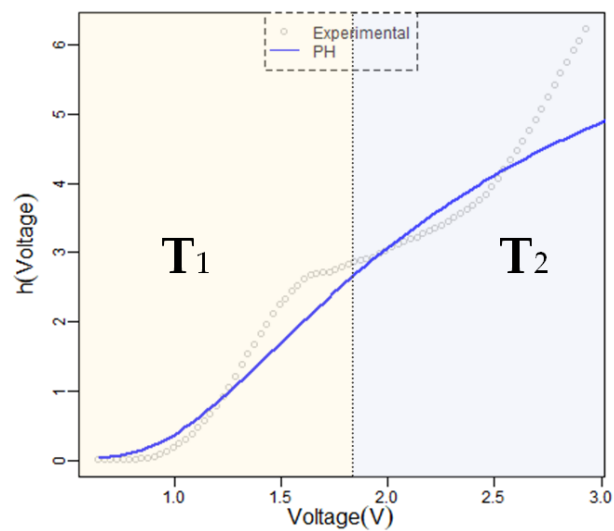

**Figure 2**: Statistical PH fits of the Hazard Rate Function of reset voltages considered in [11].

The probability distribution for this new approach can be built as follows. The reliability function will depend on whether the survival time occurs earlier or later than $a$.

The variables $X_1$ and $X_2$ are defined as the event time when occurs before than $a$ and the remaining time up to the event from $a$ when it occurs after $a$, respectively. Therefore, $X = X_1$ for $0 \leq x \leq a$ and $X = a + X_2$ for $x > a$.

- If $x \leq a$, the reliability function for $X$ is the same than the reliability function for $X_1$, and it follows a phase type distribution with representation $(\boldsymbol{\alpha}, \mathbf{T}_1)$ in this period of time. Then,

$$R(x) = P(X > x) = P(X_1 > x) = \boldsymbol{\alpha} e^{\mathbf{T}_1 x}\mathbf{e}.$$

- The variable $X_2$, remaining time up to the event from $a$, follows a phase type distribution with representation $(\boldsymbol{\alpha} e^{\mathbf{T}_1 a}, \mathbf{T}_2)$. Then, if the event occurs after $a$, that is, $x > a$,

$$R(x) = P(X > x) = P(a + X_2 > x)$$
$$= P(X_2 > x - a) = \boldsymbol{\alpha} e^{\mathbf{T}_1 a} e^{\mathbf{T}_2 (x-a)}\mathbf{e}.$$

This last scenario can be interpreted as follows: at time $a$ the event does not occur and the initial distribution for this second period of time is $\boldsymbol{\alpha} e^{\mathbf{T}_1 a}$, then the probability of failure at a certain time after $a$ is governed by matrix $\mathbf{T}_2$.

## 3. The one cut-point PH distribution

The one cut-point distribution is defined as follows.

**Definition**. A variable $X$, defined in the non-negative real semi line, is a one cut-point phase type distribution with representation $(a, \boldsymbol{\alpha}, \mathbf{T}_1, \mathbf{T}_2)$ if its probability density function is given by

$$f(x) = \begin{cases} \boldsymbol{\alpha} e^{\mathbf{T}_1 x} \mathbf{T}_1^0 & ; \quad x \leq a \\ \boldsymbol{\alpha} e^{\mathbf{T}_1 x} e^{\mathbf{T}_2 (x-a)} \mathbf{T}_2^0 & ; \quad x > a. \end{cases}$$

From the definition, the cumulative distribution function is worked out,

$$F(x) = P(X \leq x) = \begin{cases} \int_0^x \boldsymbol{\alpha} e^{\mathbf{T}_1 u}\mathbf{T}_1^0 du = 1 - \boldsymbol{\alpha} e^{\mathbf{T}_1 x}\mathbf{e} & ; \quad x \leq a \\ \int_0^a \boldsymbol{\alpha} e^{\mathbf{T}_1 u}\mathbf{T}_1^0 du + \int_a^x \boldsymbol{\alpha} e^{\mathbf{T}_1 a} e^{\mathbf{T}_2 (u-a)} \mathbf{T}_2^0 du = 1 - \boldsymbol{\alpha} e^{\mathbf{T}_1 a} e^{\mathbf{T}_2 (x-a)}\mathbf{e} & ; \quad x > a, \end{cases}$$

and therefore the reliability function is

$$R(x) = P(X > x) = \begin{cases} \boldsymbol{\alpha} e^{\mathbf{T}_1 x}\mathbf{e} & ; \quad x \leq a \\ \boldsymbol{\alpha} e^{\mathbf{T}_1 a} e^{\mathbf{T}_2 (x-a)}\mathbf{e} & ; \quad x > a, \end{cases}$$

as it was shown in previous section.

Two well-known interesting measures in the reliability field are the hazard rate and the cumulative hazard rate functions whose expressions are given, respectively, as

$$h(x) = \frac{f(x)}{R(x)} = \begin{cases} \dfrac{\boldsymbol{\alpha} e^{\mathbf{T}_1 x}\mathbf{T}_1^0}{\boldsymbol{\alpha} e^{\mathbf{T}_1 x}\mathbf{e}} & ; \quad x \leq a \\ \dfrac{\boldsymbol{\alpha} e^{\mathbf{T}_1 x} e^{\mathbf{T}_2 (x-a)}\mathbf{T}_2^0}{\boldsymbol{\alpha} e^{\mathbf{T}_1 a} e^{\mathbf{T}_2 (x-a)}\mathbf{e}} & ; \quad x > a, \end{cases}$$

and

$$H(x) = \begin{cases} -\ln\left(\boldsymbol{\alpha} e^{\mathbf{T}_1 x}\mathbf{e}\right) & ; \quad x \leq a \\ -\ln\left(\boldsymbol{\alpha} e^{\mathbf{T}_1 a} e^{\mathbf{T}_2 (x-a)}\mathbf{e}\right) & ; \quad x > a. \end{cases}$$

It is interesting to highlight that the probability density function and the hazard rate function are not continuous at point $a$, however the reliability function and the cumulative hazard rate function do verify this property.

The characteristic function provides an alternative way for describing a random variable. It is defined as $\varphi_X(t) = E\left[e^{itX}\right] = \int_{-\infty}^{\infty} e^{itx} f(x) dx$. From the one cut-point density function, the characteristic function has the following matrix-algorithmic expression for $t$ in a neighbourhood of zero,

$$\varphi_X(t) = \int_0^\infty e^{itx} f(x) dx$$

$$= \int_0^a e^{itx} \boldsymbol{\alpha} e^{\mathbf{T}_1 x} \mathbf{T}_1^0 dx + \int_a^\infty e^{itx} \boldsymbol{\alpha} e^{\mathbf{T}_1 a} e^{\mathbf{T}_2(x-a)} \mathbf{T}_2^0 dx$$

$$= \boldsymbol{\alpha} \int_0^a e^{x(\mathbf{T}_1 + it\mathbf{I})} dx \mathbf{T}_1^0 + \boldsymbol{\alpha} e^{\mathbf{T}_1 a} e^{-\mathbf{T}_2 a} \int_a^\infty e^{x(\mathbf{T}_2 + it\mathbf{I})} dx \mathbf{T}_2^0$$

$$= \boldsymbol{\alpha} \left( e^{a(\mathbf{T}_1 + it\mathbf{I})} - \mathbf{I} \right) (\mathbf{T}_1 + it\mathbf{I})^{-1} \mathbf{T}_1^0 - \boldsymbol{\alpha} e^{\mathbf{T}_1 a} e^{-\mathbf{T}_2 a} e^{a(\mathbf{T}_2 + it\mathbf{I})} (\mathbf{T}_2 + it\mathbf{I})^{-1} \mathbf{T}_2^0$$

$$= \boldsymbol{\alpha} \left( e^{a(\mathbf{T}_1 + it\mathbf{I})} - \mathbf{I} \right) (\mathbf{T}_1 + it\mathbf{I})^{-1} \mathbf{T}_1^0 - \boldsymbol{\alpha} e^{\mathbf{T}_1 a} e^{ait\mathbf{I}} (\mathbf{T}_2 + it\mathbf{I})^{-1} \mathbf{T}_2^0$$

$$= \boldsymbol{\alpha} \left( e^{a(\mathbf{T}_1 + it\mathbf{I})} - \mathbf{I} \right) (\mathbf{T}_1 + it\mathbf{I})^{-1} \mathbf{T}_1^0 - \boldsymbol{\alpha} e^{a(\mathbf{T}_1 + it\mathbf{I})} (\mathbf{T}_2 + it\mathbf{I})^{-1} \mathbf{T}_2^0$$

$$= \boldsymbol{\alpha} e^{a(\mathbf{T}_1 + it\mathbf{I})} \left[ (\mathbf{T}_1 + it\mathbf{I})^{-1} \mathbf{T}_1^0 - (\mathbf{T}_2 + it\mathbf{I})^{-1} \mathbf{T}_2^0 \right] - \boldsymbol{\alpha} (\mathbf{T}_1 + it\mathbf{I})^{-1} \mathbf{T}_1^0.$$

Following a similar reasoning, the moment-generating function is worked out

$$M_X(t) = E\left[ e^{tX} \right] = \int_0^\infty e^{tx} f(x) dx$$

$$= \boldsymbol{\alpha} e^{a(\mathbf{T}_1 + t\mathbf{I})} \left[ (\mathbf{T}_1 + t\mathbf{I})^{-1} \mathbf{T}_1^0 - (\mathbf{T}_2 + t\mathbf{I})^{-1} \mathbf{T}_2^0 \right] - \boldsymbol{\alpha} (\mathbf{T}_1 + t\mathbf{I})^{-1} \mathbf{T}_1^0.$$

Since $\left. \frac{\partial^n \varphi_X(t)}{\partial t^n} \right|_{t=0} = i^n E\left[ X^n \right]$, the first two moments have been worked out. Thus, for the mean we have

$$\frac{\partial \varphi_X(t)}{\partial t} = ai\boldsymbol{\alpha} e^{a(\mathbf{T}_1 + it\mathbf{I})} \left[ (\mathbf{T}_1 + it\mathbf{I})^{-1} \mathbf{T}_1^0 - (\mathbf{T}_2 + it\mathbf{I})^{-1} \mathbf{T}_2^0 \right]$$

$$+ i\boldsymbol{\alpha} e^{a(\mathbf{T}_1 + it\mathbf{I})} \left[ -(\mathbf{T}_1 + it\mathbf{I})^{-2} \mathbf{T}_1^0 + (\mathbf{T}_2 + it\mathbf{I})^{-2} \mathbf{T}_2^0 \right] + i\boldsymbol{\alpha} (\mathbf{T}_1 + it\mathbf{I})^{-2} \mathbf{T}_1^0.$$

Given that $-\mathbf{T}_i \mathbf{e} = \mathbf{T}_i^0$ and therefore $-\mathbf{e} = \mathbf{T}_i^{-1} \mathbf{T}_i^0$ and $-\mathbf{T}_i^{-1} \mathbf{e} = \mathbf{T}_i^{-2} \mathbf{T}_i^0$, then

$$\left. \frac{\partial \varphi_X(t)}{\partial t} \right|_{t=0} = i\boldsymbol{\alpha} e^{a\mathbf{T}_1} \left[ \mathbf{T}_1^{-1} \mathbf{e} - \mathbf{T}_2^{-1} \mathbf{e} \right] - i\boldsymbol{\alpha} \mathbf{T}_1^{-1} \mathbf{e} = iE[X].$$

The matrix expressions for the moments are

$$\mu = E[X] = -\boldsymbol{\alpha} \mathbf{T}_1^{-1} \mathbf{e} + \boldsymbol{\alpha} e^{\mathbf{T}_1 a} \left( \mathbf{T}_1^{-1} - \mathbf{T}_2^{-1} \right) \mathbf{e},$$

$$E\left[ X^2 \right] = 2\boldsymbol{\alpha} \mathbf{T}_1^{-2} \mathbf{e} - 2\boldsymbol{\alpha} e^{\mathbf{T}_1 a} \left[ \mathbf{T}_2^{-1} \left( a\mathbf{I} - \mathbf{T}_2^{-1} \right) - \mathbf{T}_1^{-1} \left( a\mathbf{I} - \mathbf{T}_1^{-1} \right) \right] \mathbf{e}.$$

**Parameter estimation**

Parameter estimation of one cut-point PH distributions refers to finding a representation $(a, \boldsymbol{\alpha}, \mathbf{T}_1, \mathbf{T}_2)$ of order $n$ from a sample size equal to $m$, $\{x_1, x_2, ..., x_m\}$. The sample points is a realization of $\{X_1, X_2, ..., X_m\}$, independent and identically distributed random variables. We will estimate the parameters by considering the well-known maximum likelihood method.

Each observation of the sample, $x_i$, contributes to the likelihood function with the probability density function, that is,

- If $x_i$ is less or equal than the cut-point $a$, then the contribution is

$$f(x_i) = \boldsymbol{\alpha} e^{\mathbf{T}_1 x_i} \mathbf{T}_1^0.$$

- If $x_i$ is greater than $a$, then at time $a$ the transient distribution of the internal Markov process is $\boldsymbol{\alpha} e^{\mathbf{T}_1 a}$ and it survives a time $t-a$ in the second period. Then, it contributes with

$$f(x_i) = \boldsymbol{\alpha} e^{\mathbf{T}_1 a} e^{\mathbf{T}_2(x_i - a)} \mathbf{T}_2^0.$$

Then, the likelihood function is given by

$$L(a, \boldsymbol{\alpha}, \mathbf{T}_1, \mathbf{T}_2) = \prod_i \left( \boldsymbol{\alpha} e^{\mathbf{T}_1 x_i} \mathbf{T}_1^0 \right)^{I_{\{x_i \leq a\}}} \left( \boldsymbol{\alpha} e^{\mathbf{T}_1 a} e^{\mathbf{T}_2(x_i - a)} \mathbf{T}_2^0 \right)^{I_{\{x_i > a\}}},$$

where $I_{\{\}}$ is the indicator function.

Therefore the log-likelihood function is given by

$$\log L(a, \boldsymbol{\alpha}, \mathbf{T}_1, \mathbf{T}_2) = \sum_{x_i \leq a} \log\left( \boldsymbol{\alpha} e^{\mathbf{T}_1 x_i} \mathbf{T}_1^0 \right) + \sum_{x_i > a} \log\left( \boldsymbol{\alpha} e^{\mathbf{T}_1 a} e^{\mathbf{T}_2(x-a)} \mathbf{T}_2^0 \right).$$

The likelihood function and the results shown in this section have been computationally implemented in R-cran.

## 4. Application

The conductive filament creation and rupture (i.e. the resistive switching process) in Resistive Random Access Memories according to several characteristics is analysed by applying one cut-point PH distributions. The CF creation is an inherent stochastic process since different ions in the dielectric are moved by the electric field and influenced by temperature in a random manner [23]. The ions form clusters that finally configure a percolation path that constitutes the CF. Since the generation of the ions and their movement is random, the whole CF evolution is therefore inherently random. This leads to the known cycle-to-cycle (C2C) variability [12] i.e, a variability that reflects that the device resistance changes each time the CF is formed, in each cycle. In addition to the C2C variability, RRAM device-to-device variability due to differences between devices in the fabrication process of a chip can also be seen [12]. The former, the C2C variability, is the subject of our study here.

We consider devices of a 10nm thick $HfO_2$ dielectric in between a Ti and W electrodes [24], see the inset in Figure 3a. The different current-voltage curves (Figure 3a) have been measured using low slope voltages ramps with time. The set and reset voltages and currents (Figure 3b) have been determined for a long series of 1000 resistive switching cycles. Figures 3c and 3d are typical plots that show C2C variability.

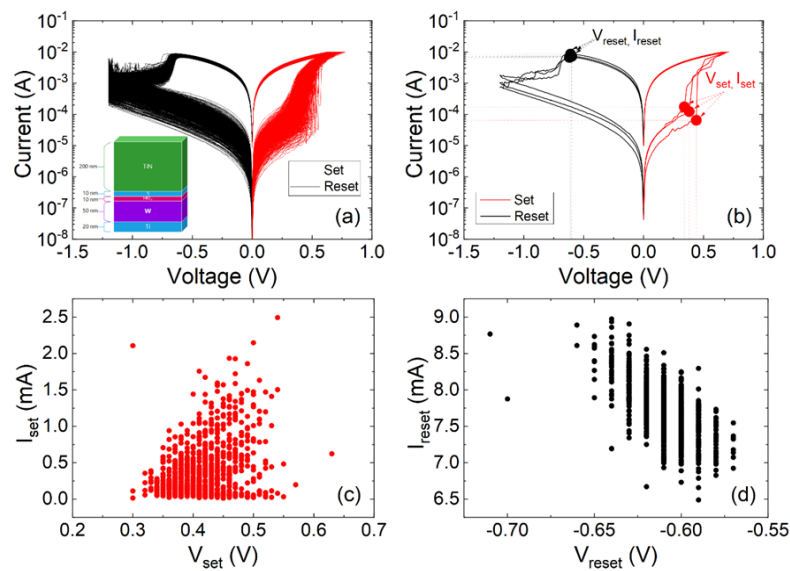

**Figure 3**: a) Experimental I-V curves for 1000 set/reset consecutive cycles measured. The layer stack scheme for the devices under study is shown in the inset, b) experimental I-V curves for 3 set/reset cycles showing Vset, Vreset, Iset and Ireset, c) experimental Iset versus Vset data, d) experimental Ireset versus Vreset data.

Some of the resistive switching parameters (RSP) that can be employed to characterize a RRAM technology are the set and reset voltages and currents, see Figure 3b. These RSP identify the voltage and current levels that should be used in circuits, in which RRAMs are employed as components. In this respect, the correct modeling of the behaviour would allow circuit designers to determine the operation limits of their circuits, that is why this issue is essential. In addition to the direct analysis of the values obtained, as it is presented by most of experimental works, the advanced statistical modeling of the RSP data, permits a better variability study from the circuit simulation viewpoint. Different complex approaches have been introduced to model these data using a unimodal statistical distribution [11, 25]. While it is true that these complex models get better results than the classical methodologies applied in the electronics industry until now, they might not be accurate enough in certain occasions. On account of this weakness, we apply the new methodology to describe the experimental data. In this respect, as it will be shown below, a better representation of the distribution tails is achieved and the number of parameters to be estimated is reduced as well. This analysis has been carried out with R-cran by using the implementation developed for this objective. Regarding the process of parameters optimization, the "*optim*" function available in R-cran has been used through L-BFGS-B method [26]. This method uses a limited-memory modification of the quasi-Newton method and enables to stablish a lower and/or upper bound for the parameters.

*The reset voltage*

The reset voltage describes the voltage level at which the conductive filament is broken and the resistance of the devices increases greatly. To analyse the reset voltage, a data set with 1000 experimental observations has been considered. The empirical voltage mean and the standard deviation are 0.6091 and 0.0158, respectively. Firstly, a PHD has been fitted to the data set by considering a general internal structure. After analysis, fixing the number of phases, all the results converge to the same internal structure, which corresponds an Erlang distribution with representation PH. This structure can be expressed as follows:

$$\boldsymbol{\alpha} = (1,0,...,0) \text{ and } \mathbf{T} = \begin{pmatrix} -\lambda & \lambda & 0 & \cdots & 0 \\ 0 & -\lambda & \lambda & \ddots & 0 \\ \vdots & & \ddots & \ddots & 0 \\ 0 & 0 & \cdots & -\lambda & \lambda \\ 0 & 0 & \cdots & 0 & -\lambda \end{pmatrix}_{n \times n}.$$

For this dataset, even if the number of states is too high, the goodness of fit was analyzed from the Anderson-Darling (A-D) test, obtaining a p-value less than 0.0001 and rejecting the goodness of the fit. Hereinafter, and for computational reasons, a PH with the structure described above with 200 phases is considered.

A one cut-point defined in Section 3 has been considered to estimate the reset voltage. To simplify the model, a one cut-point PH distribution with representation $(a, \boldsymbol{\alpha}, \mathbf{T}_1, \mathbf{T}_2)$, where $\mathbf{T}_1$ and $\mathbf{T}_2$ have Erlang internal structure, has been assumed. In total 4 parameters were estimated, the cut-point is $a = 0.595$ (confidence interval [0.571; 0.619] at 95%), the number of phases is reduced considerably (only 14 phases are needed), and the parameters of $\mathbf{T}_1$ and $\mathbf{T}_2$ are,

$$\mathbf{T}_1 = \begin{pmatrix} -\lambda_1 & \lambda_1 & 0 & \cdots & 0 \\ 0 & -\lambda_1 & \lambda_1 & \ddots & 0 \\ \vdots & & \ddots & \ddots & 0 \\ 0 & 0 & \cdots & -\lambda_1 & \lambda_1 \\ 0 & 0 & \cdots & 0 & -\lambda_1 \end{pmatrix}_{14 \times 14} ; \mathbf{T}_2 = \begin{pmatrix} -\lambda_2 & \lambda_2 & 0 & \cdots & 0 \\ 0 & -\lambda_2 & \lambda_2 & \ddots & 0 \\ \vdots & & \ddots & \ddots & 0 \\ 0 & 0 & \cdots & -\lambda_2 & \lambda_2 \\ 0 & 0 & \cdots & 0 & -\lambda_2 \end{pmatrix}_{14 \times 14}$$

with $\lambda_1 = 16.74531$ and $\lambda_2 = 261.61844$. Vector $\boldsymbol{\alpha}$ continues being $\boldsymbol{\alpha} = (1,0,...,0)$.

Figure 4 shows the cumulative hazard rate and the probability density function for the empirical, PH and cut-point PH cases. Note that the empirical density has been worked out by means of "*density*" function available in R-cran. By default, this function disperses the mass of the empirical distribution over a regular grid of points and makes use of the Fourier transform in order to combine this approximation with a discretized version of the kernel, so that to use linear approximation to evaluate the density at the specified points [27-28]. The cut-point approach not only reduces the parameters to be estimated, but also improves the quality of the fitting. Table 1 shows the main empirical characteristics and a comparison between models.

Table 1. Comparing measures between models for reset voltage.

|  | Mean | Standard deviation | Number phases | Estimates | A-D test |
|---|---|---|---|---|---|
| **Empirical** | 0.6091 | 0.0158 | | | |
| **PH model** | 0.6091 | 0.0431 | 200 | $\lambda = 164.1767$ | 0.000 |
| **Cut-point** | 0.6003 | 0.0431 | 14 | $\lambda_1 = 16.74531$<br>$\lambda_2 = 261.61844$ | 0.023 |

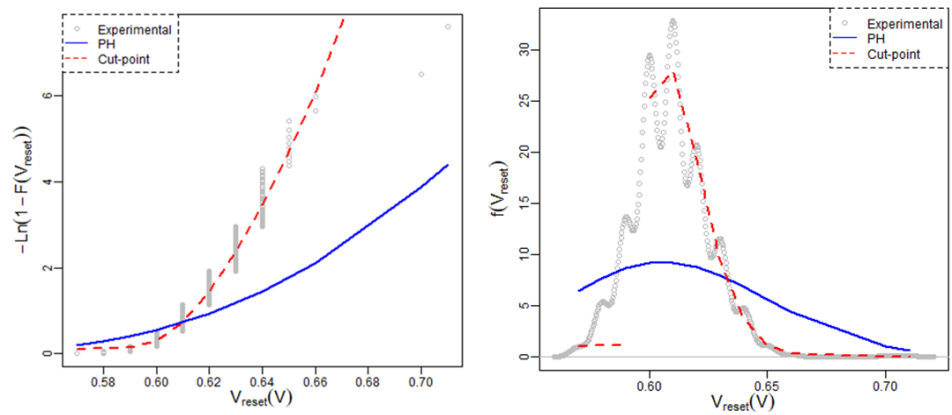

**Figure 4**: Cumulative hazard rate (left) and probability density function (right) for 1000 reset voltages (the absolute value was employed for the plot) and the corresponding PH and cut-point PH fit.

*Reset current*

The reset current represents the current at the reset point and it is statistically linked to the CF size at the point where it is ruptured. There is a connection between this parameter and the reset voltage; however, different factors can statistically decouple them, as can be clearly seen in Figure 3d. For this parameter, a multimodal distribution analysis is expected to work also better.

A cut-point PH distribution has been fitted for the reset current and it has been compared with a PH adjustment. This analysis has been carried out from the sample we are employing, whose size is equal to 1000. Analogously to the reset voltage, the internal structure for the PH and one cut-point PH distribution is Erlang. In this case, the estimated cut-point is equal to *a*= 0.0072 with confidence interval [0.0068; 0.0076] at 95%.   Table 2 shows a summary of the models, adjustment and estimates.

As it can be seen in Table 2, the number of phases decreases considerably and the Anderson-Darling test rejects the PH model but not the one cut-point model. Figure 5 shows the fitting graphically.

**Table 2.** Summary for the PH and one cut-points adjustment for the reset current.

|  | Mean | Standard deviation | Number phases | Estimates | A-D test |
|---|---|---|---|---|---|
| **Empirical** | 0.0077 | 0.0004 |  |  |  |
| **PH model** | 0.0077 | 0.0004 | 353 | $\lambda$ = 45888.49 | 0.003 |
| **Cut-point** | 0.0076 | 0.0005 | 12 | $\lambda_1$ = 1003.27<br>$\lambda_2$ = 9652.37 | 0.141 |

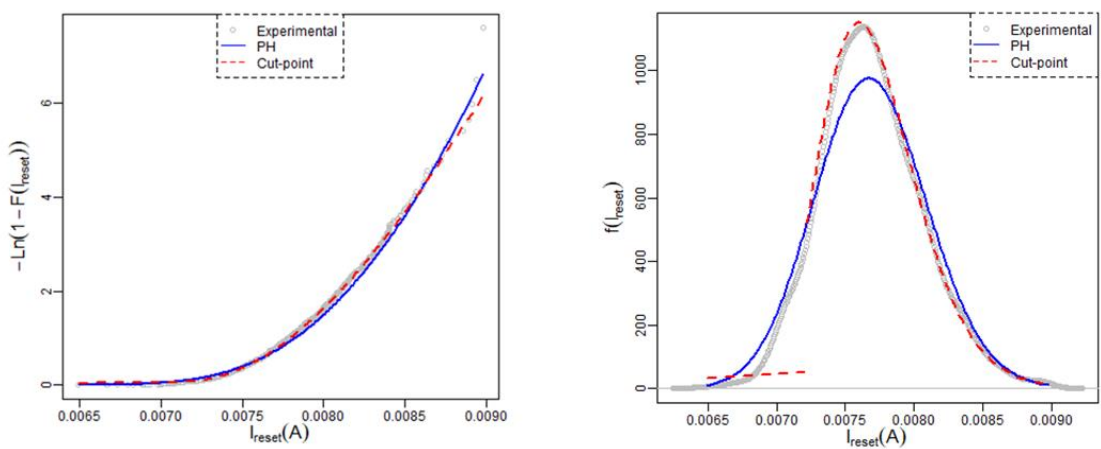

**Figure 5**: Cumulative hazard rate (left) and probability density function (right) for 1000 reset current values, and the corresponding PH and cut-point PH fit.

*The set voltages*

The set voltage marks the conduction point in which the percolation path that constitutes the conductive filament is fully formed. The randomness connected to the physical mechanisms involved in the CF formation is reflected in the experimental data variability. We also present here a multimodal approach to tackle the set voltage analysis.

A similar study has been performed for the time up to the set voltage. A sample with 1000 elements is taken into account corresponding to the same resistive switching series. The estimated cut-point is $a$ = 0.315 with confidence interval [0.296; 0.334] at 95%. The empirical results and the data for the PH and the one cut-point PH fit is shown in Table 3.

Table 3. Summary for the PH and one cut-point adjustment for the set voltage.

|  | Mean | Standard deviation | Number phases | Estimates | A-D test |
|---|---|---|---|---|---|
| **Empirical** | 0.4147 | 0.0445 |  |  |  |
| **PH model** | 0.4147 | 0.0440 | 89 | $\lambda$ = 214.6181 | 0.0147 |
| **Cut-point** | 0.4147 | 0.0451 | 11 | $\lambda_1$ = 11.5570 $\lambda_2$ = 73.7963 | 0.0571 |

The fitting is shown in Figure 6. Here we can see how the cut-point fit rectifies the lack of precision in distribution tail.

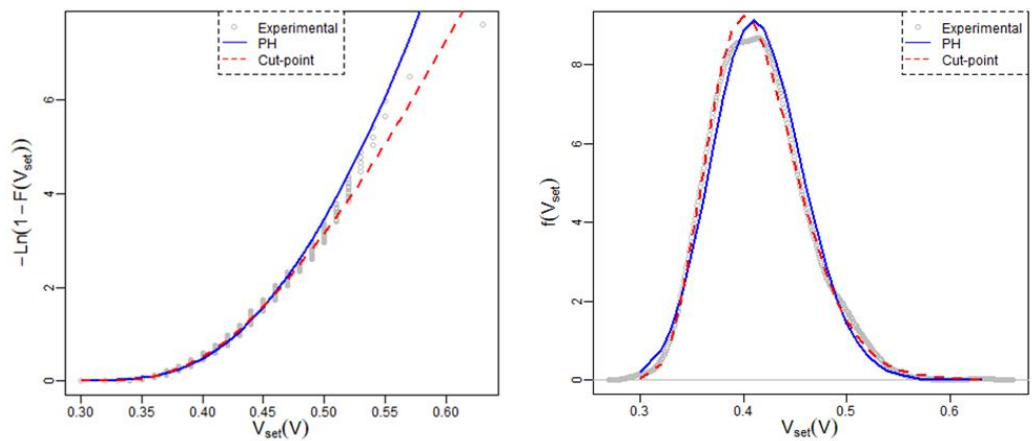

**Figure 6**: Cumulative hazard rate (left) and probability density function (right) for 1000 set voltages and the corresponding PH and cut-point PH fit.

*The set current*

The set current is the current level at the set voltage in the current-voltage curve. The current level is much lower than the reset current since the device start the set process in a High Resistance State, so, the first point with the percolation path formed represents a low current state. Again, the resistive switching characteristic randomness can be observed. As highlighted in the reset current case, there are factors that affect the correlation between of the set voltage and current data, as can be seen in Figure 3c.

Table 4 shows the obtained results after performing an exhaustive analysis on the behaviour of the set current by means of the new methodology presented in the current manuscript. The estimated cut point is $a$ = 0.00025 with confidence interval [0.00019; 0.00031] at 95%.

The new model of one cut-point improves the fit to the data set, without rejecting the case of the new model. A graphical analysis is shown in Figure 7.

Table 4. Summary for the PH and one cut-point adjustment for the set current.

|  | Mean | Standard deviation | Number phases | Estimates | A-D test |
|---|---|---|---|---|---|
| **Empirical** | 0.0004 | 0.0004 |  |  |  |
| **PH model** | 0.0004 | 0.0004 | 1 | $\lambda$ = 2560.425 | 0.0001 |
| **Cut-point** | 0.0004 | 0.0004 | 2 | $\lambda_1$ = 6820.583 $\lambda_2$ = 3495.02 | 0.0819 |

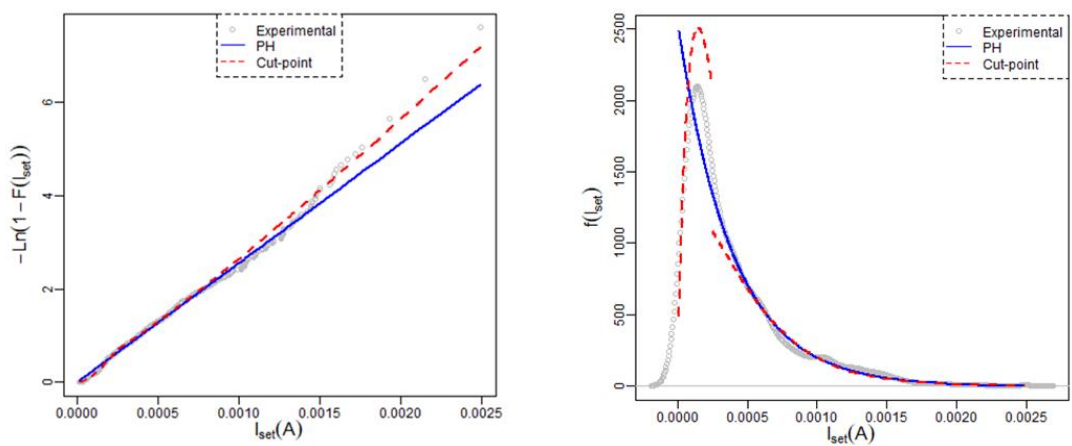

**Figure 7**: Cumulative hazard rate (left) and probability density function (right) for 1000 set current and the corresponding PH and cut-point PH fit.

## 5. Conclusions

The good properties of phase-type distributions make this class of distributions a suitable candidate to model experimental data in the field of reliability. Among other features, PH enables the interpretation of the results in a simple way thanks to its matrix-algebraic form and moreover, they generalize a large number of known distributions such as Exponential, Erlang or Coxian distribution. In addition, a reason why this class is usually considered in many applications is due to the fact that it is dense in the set of probability distributions defined on any half-line of real number. This assertion implies that any non-negative probability distributions can be approximated as much as desired by a PH. The PH inherent problem is that their fittings show certain weaknesses in several aspects.

Sometimes the adjustment is improvable in the tails of the distribution (especially in heavy distributions) and in other occasions, even for an appropriate fitting, the number of parameters to estimate is really high. In this respect, a new class of distributions, called one cut-point PH distributions, is introduced in the current work in order to solve the problems aforementioned. This distribution, which is a first approach to non-homogeneous PH distributions, inherits the most important characteristics of the original PH.

The motivation of this paper is to provide a new tool to model experimental data measured on Resistive Random Access Memories, one of the most promising devices in the current semiconductor industry landscape. The results presented show that the fitting considering a cut-point PH distribution improves significantly the obtained results in comparison to the classic PH.

**Author Contributions:** J.E.R. and C.A. developed all the statistical theory, computationally implemented the methodology, and obtained the results shown. J.B.R. and D.M. were in charge of the physical theoretical explanation and its conclusions, and they performed the data curation for this study. All authors contributed equally to the writing part of the manuscript. All authors have read and agreed to the published version of the manuscript.

**Funding:** This paper is partially supported by the project FQM-307 of the Government of Andalusia (Spain), by the project PID2020-113961GB-I00 of the Spanish Ministry of Science and Innovation (also supported by the European Regional Development Fund program, ERDF) and by the project PPJIB2020-01 of the University of Granada. Also, the first and second authors acknowledge financial support by the IMAG–María de Maeztu grant CEX2020-001105-M / AEI / 10.13039/501100011033. They also acknowledge the financial support of the Consejería de Conocimiento, Investigación y Universidad, Junta de Andalucía (Spain) and the FEDER programme for projects A.TIC.117.UGR18 and IE2017-5414.

**Data Availability Statement**: The data sets generated and/or analysed during the current study are available from the corresponding author on reasonable request.

**Acknowledgments:** We would like to thank F. Campabadal and M. B. González from the IMB-CNM (CSIC) in Barcelona for fabricating and measuring the devices employed here.

**Conflicts of Interest:** The authors declare no conflict of interest. The funders had no role in the design of the study; in the collection, analyses, or interpretation of data; in the writing of the manuscript, or in the decision to publish the results.

**Abbreviations**

The following abbreviations are used in this manuscript:

| | |
|---|---|
| C2C | Cycle-to-cycle |
| CF | Conductive Filament |
| HfAlO | Oxide Al-doped HfO2 |
| PH | Phase-type |
| RRAM | Resistive Random Access Memory |
| RSP | Resistive switching parameters |